\documentclass[aps]{revtex4}
\usepackage{graphicx}

\begin{document}

\title{Scattering solutions of the spinless Salpeter equation}
\author{F. Brau\thanks{Chercheur I.I.S.N.} 
and C. Semay\thanks{Chercheur qualifi\'{e} F.N.R.S.}}
\affiliation{Groupe de Physique Nucl\'eaire Th\'eorique, Universit\'e
de Mons-Hainaut, B-7000 Mons, Belgium}
\date{\today}

\begin{abstract}
A method to compute the scattering solutions of 
a spinless Salpeter equation (or a Schr\"{o}dinger equation) with
a central interaction
is presented. This method relies on the 3-dimensional 
Fourier grid Hamiltonian method used to compute bound states. 
It requires only the evaluation of the potential
at equally spaced grid points and yields the radial part of the 
scattering solution 
at the same grid points. It can be easily extended to the 
case of coupled channel equations and to the case of non-local
interactions.
\end{abstract}
 % PACS
% Other computational methods
% Relativistic wave equations
% Relativistic scattering theory

\maketitle

\section{Introduction}
\label{sec:intro}

Numerous techniques have been developed to compute the scattering 
solutions of a Schr\"{o}dinger equation. Simple Runge-Kutta methods
can be performed in the case of a local potential, and discretization
of the integration domain can be used for a non-local interaction
\cite{pres92}. All these techniques can be used because the
kinetic energy operator can be expressed in terms of a
derivative operator. This is no longer true in the case of a spinless
Salpeter equation for which the kinetic energy is a complicated
square-root operator. 

\par In a previous paper \cite{brau97a}, we have developed a method
to compute the eigenvalues of a spinless Salpeter equation. 
This method relies on the fact that the kinetic 
energy operator is best represented in momentum space, while the
potential energy is generally given in coordinate space.
It requires only the evaluation of the potential
at equally spaced grid points, and yields directly the amplitude of
the solution at the same grid points. This method is derived
from the Fourier grid Hamiltonian method \cite{mars89,bali91}
developed to compute the solution of the one-dimensional
Schr\"{o}dinger equation, and consequently was called the
3-dimensional Fourier grid Hamiltonian method. It appears very 
accurate and simple to handle.  

\par In this paper, we show that 3-dimensional Fourier grid
Hamiltonian method can be used to compute the scattering solutions of
a spinless Salpeter equation (or a Schr\"{o}dinger equation). We
focus our attention on the case of a purely central local potential,
but the method can also be applied if the potential is non-local, or
if couplings exist between different channels. Up to our knowledge,
this is the first time that the scattering solutions of the spinless
Salpeter equation are presented.

\par Our method is outlined in Sec.~\ref{sec:method}.
Test applications of the method are presented in Sec.~\ref{sec:num},
and a brief summary is given in Sec.~\ref{sec:summary}.

\section{Method}
\label{sec:method}

\subsection{Theory}
\label{sec:theory}

We assume
that the Hamiltonian can be written as the sum of the kinetic
energy $\hat T$ and a potential energy operator $\hat V$. The 
scattering equation is given by
\begin{equation}
\label{tv}
\left[ \hat T + \hat V \right] |\Psi \rangle = E |\Psi \rangle,
\end{equation}
where $\hat T$ depends only on the square of the relative impulsion 
$\vec p$ between
the particles, $\hat V$ is a local interaction which depends on
the relative distance,
and $E$ is the asymptotic kinetic energy of the two interacting
particles. 
This equation is a spinless Salpeter equation if 
\begin{equation}
\label{salp}
\hat T = \sqrt{\vec p\,^2 + m_1^2} + \sqrt{\vec p\,^2 + m_2^2} - m_1
- m_2,
\end{equation}
where $m_1$ and $m_2$ are the masses of the particles
(we use the natural units $\hbar = c =1$ throughout the text). 
Equation~(\ref{tv}) is a
Schr\"{o}dinger equation if 
\begin{equation}
\label{schro}
\hat T = \frac{\vec p\,^2}{2 \mu} \qquad \text{with} \qquad 
\mu = \frac{m_1 m_2}{m_1 + m_2}.
\end{equation}
 
\par In configuration space, Eq.~(\ref{tv}) is written
\begin{equation}
\label{tvconf}
\int \left[ \langle \vec r\, |\hat T |\vec r\,' \rangle +
\langle \vec r\, |\hat V |\vec r\,' \rangle \right] 
\langle \vec r\,' |\Psi \rangle\, d\vec r\,' 
= E\, \langle \vec r\, |\Psi \rangle.
\end{equation}
In the following, we only consider the case of a local central
potential 
\begin{equation}
\label{vcent}
\langle \vec r\, |\hat V |\vec r\,' \rangle =
V(r)\, \delta(\vec r\, - \vec r\,') \quad
\text{with} \quad r=|\vec r\,|.
\end{equation}
Consequently, the wave function has the following form
\begin{equation}
\label{focent}
\langle \vec r\, |\Psi \rangle = R_l(r)\, Y_{lm}(\hat r) \quad
\text{with} \quad \hat r = \vec r / r.
\end{equation}
Using the method developed in Ref.~\cite{brau97a},
Eq.~(\ref{tvconf}) can be rewritten as
\begin{equation}
\label{eqrad}
\frac{2}{\pi}\, r \int_0^\infty dr'\, r'\, u_l(r') \int_0^\infty dq\,
q^2\,
T(q^2)\, j_l(qr)\, j_l(qr') + V(r)\, u_l(r) = E\, u_l(r).
\end{equation}
where $u_l(r) = r R_l(r)$ is the regularized radial function and
where functions $j_l(qr)$ are spherical Bessel functions. 

\par Using the following orthogonality relation 
\begin{equation}
\label{jj}
\frac{2}{\pi}\, x\,x' \int_0^\infty j_l(qx)\, j_l(qx')\, q^2\, dq =
\delta(x-x').
\end{equation}
One can show that $u_l(r) \propto r j_l(kr)$, with $k$ fixed, is a
solution of 
Eq.~(\ref{eqrad}) with vanishing potential. The relative energy $E$
is then equal to $\sqrt{k^2 + m_1^2} + \sqrt{k^2 + m_2^2} -
m_1 - m_2$ in the case of a spinless Salpeter equation, and is equal to
$k^2/(2 \mu)$ in the case of a Schr\"{o}dinger
equation.

\subsection{Discretization}
\label{sec:discret}

In order to compute the scattering solutions of Eq.~(\ref{eqrad}), we
replace the continuous variable $r$ by a grid of discrete
values
$r_i$ defined by
\begin{equation}
\label{ri}
r_i = i\Delta \quad \text{with} \quad i=0,\, 1,\, \ldots,\, N,
\end{equation}
where $\Delta$ is the uniform spacing between the grid points. 
Regularity at origin $r_0 = 0$ imposes $u_l(r_0) = 0$. In the
following, we
always consider potential with a finite range $\lim_{r \rightarrow 
\infty} r\,V(r) = 0$ (the case of scattering by a Coulomb-like potential
is not considered here). Outside the range of the potential, the
solution
is a phase shifted free wave function. For a value of $r_N = N\Delta$
sufficiently large, we choose to set arbitrarily $u_l(r_N) = 1$ in
order to fix the normalization of the wave function. 

\par As explained in Ref.~\cite{brau97a}, the spacing $\Delta$ in the
configuration space determines the grid spacing 
$\Delta k$ in the momentum space. Therefore, 
we have a grid in the configuration space and a corresponding grid in
the momentum space
\begin{equation}
\label{gridk}
k_s = s\Delta k = \frac{s \pi}{N \Delta} 
\quad \text{with} \quad s=0,\, 1,\, \ldots,\, N.
\end{equation}
If we note $V_i = V(r_i)$, the discretization procedure
replaces the continuous Eq.~(\ref{eqrad}) by a matrix equation
\begin{equation}
\label{matrix}
\sum_{j=1}^{N-1} \left[ H_{ij} - E\, \delta_{ij} \right] \phi_j = -
H_{iN} \qquad \text{for}
\qquad i=1,\, \ldots,\, N-1,
\end{equation}
where
\begin{equation}
\label{hij}
H_{ij} = \frac{2\pi^2}{N^3}\, i\, j \sum_{s=1}^N s^2\, 
T\left( \left( \frac{\pi s}{N \Delta} \right)^2 \right)\,
j_l\left( \frac{\pi}{N}s i \right)\,
j_l\left( \frac{\pi}{N}s j \right)
+ V_i\, \delta_{ij}.
\end{equation}
The discrete solution $\phi_i$ of the linear system (\ref{matrix})
gives approximately the values of the radial part of the 
solution of Eq.~(\ref{eqrad}) at the grid points: $\phi_i \simeq
u_l(r_i)$. The phase shift can be computed by using the values of the
wave function at two points in the region where the potential is
vanishing \cite{rodb67}.

\par This method can also be used in the case of a non-local
potential and in the case of coupled channels calculations.
Some details about the implementation of such problems are given in
Ref.~\cite{brau97a}.

\par Actually, the scattering solution cannot be obtained directly
from Eq.~(\ref{matrix}). For instance, in the case of a zero angular
momentum solution, it is easy to
see that $H_{iN} = 0$. Consequently, we have to set $u_l(r_{N-1}) =
1$ and to restrict the summation in Eq.~(\ref{matrix}) to $N-2$ (the
point $u_l(r_N)$ cannot be determined). Other normalization problems
appear for all values of angular momentum. All are due to the
discretization procedure as explained below.

\par The 3-dimensional Fourier grid Hamiltonian method relies on
relation (\ref{jj}). The equivalent discrete orthogonality relation
on our grid of points is
\begin{equation}
\label{jjdisc}
\frac{2\pi^2}{N^3}\, i\, j \sum_{s=1}^N s^2\, 
j_l\left( \frac{\pi}{N}s i \right)\,
j_l\left( \frac{\pi}{N}s j \right)
= \Delta_{ij}^{(N,l)}.
\end{equation}
One can thus expect that $\Delta_{ij}^{(N,l)} = \delta_{ij}$ for all 
values of $N$ and $l$. Actually, the situation is less favorable.
In Ref.~\cite{brau97a}, we show that, for $l=0$, we have
\begin{equation}
\label{l0}
\Delta_{ij}^{(N,l=0)} = \delta_{ij} 
\qquad \text{for} \qquad i,j=1,\, \ldots,\, N-1.
\end{equation}
We have verified numerically that 
\begin{mathletters}
\label{lne0}
\begin{equation}
\label{lne0:a}
\lim_{N \rightarrow \infty} \Delta_{ij}^{(N,l=1)} = \delta_{ij} 
\qquad \text{for} \qquad i,j=1,\, \ldots,\, N-1,
\end{equation}
\begin{equation}
\label{lne0:b}
\lim_{N \rightarrow \infty} \Delta_{ij}^{(N,l>1)} \approx
\delta_{ij}
\qquad \text{for} \qquad i,j=1,\, \ldots,\, N-1.
\end{equation}
\end{mathletters}
Consequently, the accuracy of this method 
becomes poorer when $l$ increases; nevertheless for large enough
number of grid points, very good results can be obtained.

\par For scattering problems, it is also interesting to calculate the
values of $\Delta_{iN}^{(N,l)}$ quantity. One can also expect that
$\Delta_{iN}^{(N,l)} = \delta_{iN}$ for all values of $N$ and $l$.
Actually, it is easy to show that
\begin{equation}
\label{sd0}
\Delta_{iN}^{(N,l=0)} = 0.
\end{equation}
For other values of $l$, we have verified numerically that
\begin{mathletters}
\label{sdlne0}
\begin{equation}
\label{sdlne0:a}
\lim_{N \rightarrow \infty} \Delta_{iN}^{(N,l \ne 0)} = 0 
\qquad \text{for} \qquad i=1,\, \ldots,\, N 
\quad \text{and} \quad l\ \text{even},
\end{equation}
\begin{equation}
\label{sdlne0:b}
\lim_{N \rightarrow \infty} \Delta_{iN}^{(N,l \ne 0)} = 2\,\delta_{iN} 
\qquad \text{for} \qquad i=1,\, \ldots,\, N 
\quad \text{and} \quad l\ \text{odd},
\end{equation}
\end{mathletters}

\par The simple way to obtain a correct normalization for the
solutions, that is to say a value of 1 for the regularized radial
part of the wave function at the last point of integration, is to
solve two different linear systems with respect to the parity of the
angular momentum.
As we shall show in the next section, the following procedures allow
to obtain accurate solutions of the scattering problem.
\begin{mathletters}
\label{mat}
\begin{equation}
\label{mat:a}
\sum_{j=1}^{N-2} \left[ H_{ij} - E\, \delta_{ij} \right] \phi_j = -
H_{i\,N-1} \qquad \text{for}
\qquad i=1,\, \ldots,\, N-2 \quad \text{and} \quad l\ \text{even},
\end{equation}
\begin{equation}
\label{mat:b}
\sum_{j=1}^{N-1} \left[ H_{ij} - E\, \delta_{ij} \right] \phi_j = -
H_{iN}/2 \qquad \text{for}
\qquad i=1,\, \ldots,\, N-1 \quad \text{and} \quad l\ \text{odd},
\end{equation}
\end{mathletters}

\section{Numerical implementation}
\label{sec:num}

\subsection{Free solutions}
\label{sec:num1}

\begin{figure}
\centering
\includegraphics*[width=10cm]{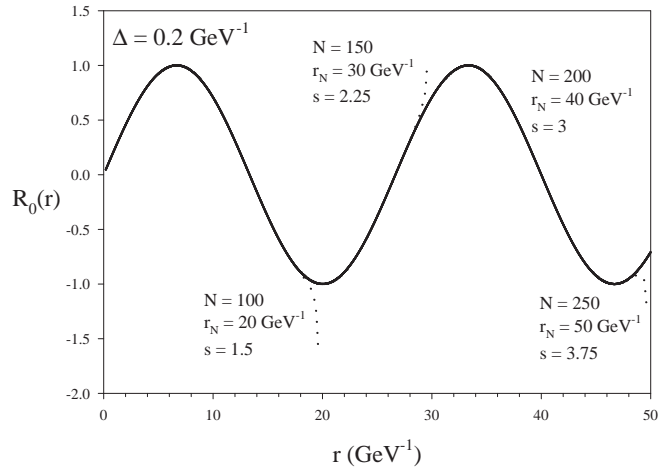}
\protect\caption{Regularized radial part $R_0(r)$ of computed
solutions
(dotted lines) and the exact solution (solid line) for the
semi-relativistic free equation with a given relative kinetic energy
and a zero angular momentum. The computed solutions are given for the
same spacing, but for different values of $r_N$. The value of the
parameter $s = N\Delta k/\pi$ is also presented. The energy is chosen
in order that $s=3$ for $r_N = 40$ GeV$^{-1}$. All computed wave
functions are normalized to match the exact solution.}
\label{fig:1}
\end{figure}

As noted above, solutions of the non-relativistic and
semi-relativistic free Eq.~(\ref{eqrad}) can be expressed in terms of
spherical Bessel functions. It can
be easily shown that the vector $\{ r_i\, j_0(k r_i); i=1,\, \ldots,\,
N-2 \}$, up to a normalization constant,
is a solution of the system (\ref{mat:a}) for $l=0$ if $k = 
\frac{\pi}{N \Delta}s$ with $s =1,\, \ldots,\, N-2$. This vector is
no longer a solution if $k$ is not a integer multiple of
$\frac{\pi}{N \Delta}$. In this case the computed solution matches
the solution of the continuous Eq.~(\ref{eqrad}) everywhere, except
near the last point $r_N$ of the domain of integration. This
situation is illustrated with Fig.~\ref{fig:1} for the
semi-relativistic free equation. The regularized radial part of the
computed solution for a given energy is presented for the same value
of the spacing, but for different values of $r_N$. The relative
kinetic energy being fixed, different values of $r_N$ correspond to
different values for the parameter $s$. We can show on this figure
that the computed solutions differ from the exact solution when $s$
is not an integer. The situation is not modified by a change of the
energy.

\par If the angular momentum $l$ is different from zero, then the
vector $\{ r_i\, j_l(k r_i) \}$ is not an exact solution of the system
(\ref{mat:a}) or (\ref{mat:b}). Nevertheless, a very good
approximation of the continuous free solution can be obtained with
correct values for the parameters $\Delta$ and $r_N$. Again, the
computed and the exact solutions can differ 
strongly near the last point $r_N$. It
is worth noting that the differences between the computed and the
exact solutions are much less large in the non-relativistic case,
whatever the value of $l$. 

\par In the free case, the phase shift is expected to be zero.
Numerically, the phase shift can be determined by using two points of
the
computed solution. If these points are chosen in the region near
$r_N$, the phase shift found can be different from zero. 
On the contrary, when the two points are taken far
from $r_N$, the phase shift value vanishes. 

\subsection{Gaussian potential}
\label{sec:num2}

We have tested our method with different finite range
potentials in the case of symmetric or asymmetric systems. In this
section, we shall only present some results obtained with a Gaussian
potential
\begin{equation}
\label{potg}
V(r) = -V_0\, e^{-r^2/a^2},
\end{equation}
for two identical particles $m_1 = m_2 = m$.

\par
\begin{figure}
\centering
\includegraphics*[width=10cm]{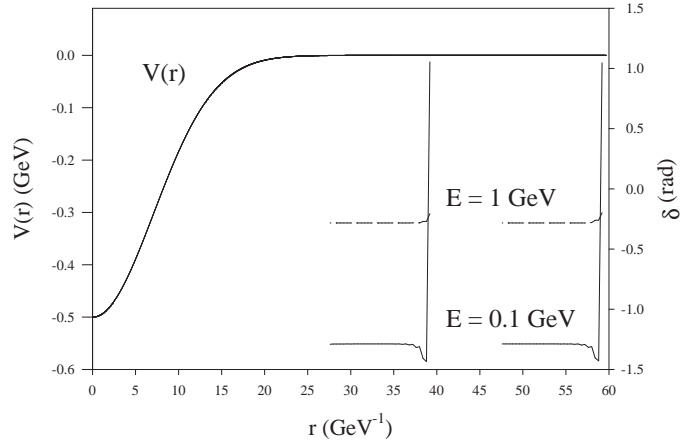}
\protect\caption{Phase shifts $\delta$ in radians for two identical
semi-relativistic particles
with $m=1$ GeV, interacting via a Gaussian potential with $V_0=0.5$
GeV and $a=10$ GeV$^{-1}$. The phase shifts are plotted as a function
of $r_p$ (see Sec.~\ref{sec:num2}) for two values of the relative energy
and for two
values of $r_N$. The potential as a function of $r$ is also
indicated.}
\label{fig:2}
\end{figure}

 In the free case, the computed solution can differ strongly from
the real solution near the last point $r_N$. This is also the case
when a potential is turned on. The phase shift can be computed with
two values of the numerical solution evaluated at two 
different points $r_p$ and $r_q$. If
$p$ or $q$ is too close to $N$, then the value of the phase shift can
be very bad. Obviously, the two points must be taken in a region where
the potential can be neglected with respect to the relative kinetic
energy. A good procedure to get reliable phase shift is to compute
the wave function with a large value of $r_N$. Then, the phase shift
can be computed with two adjacent points $r_p$ and $r_{p-1}$ as a
function of the index $p$. By decreasing the value of $p$ from $N$,
the phase
shifts will first strongly vary and rapidly reach a stable value, as
long as $r_p$ is large enough to not fall in a region where the
potential cannot be neglected. This situation is illustrated with
Fig.~\ref{fig:2}. On this figure, the phase shift for two identical
semi-relativistic particles with
$m=1$ GeV is plotted as a function of $r_p$ for two values of the
relative energy and for two values of $r_N$. The Gaussian potential 
is characterized by 
$V_0=0.5$ GeV and $a=10$ GeV$^{-1}$. It is worth noting
that the variation of the phase shifts are much larger for the
semi-relativistic case than for the non-relativistic case. 

\par Scattering states have been calculated with our method for two
non-relativistic
particles interacting with a Gaussian potential. In this case, wave
functions and phases shifts can also be computed with a great variety
of methods. For a large range of relative energy and for angular
momentum varying from 0 to 4, we have checked that
all approaches give same results.
Within our method, a relative accuracy of at least 
$10^{-4}$ for phase shifts
can be obtained with a grid containing 200--400 points. Obviously,
the interest of our method is to compute scattering solutions in the
semi-relativistic case. 

\par It is shown in the appendix that a spinless Salpeter equation
with a particular separable
non-local potential can be transformed into a Schr\"{o}dinger-like
non-local equation. In this case, the scattering semi-relativistic
equation can be solved directly by the Fourier grid Hamiltonian
method, or using the equivalent
Schr\"{o}dinger-like form, by usual techniques. We have verified, for
several values of the parameters and for different values of the
relative kinetic energy, that all methods give the same results. This
yields a direct verification of our approach. 

\par Finally, we give, in Table~\ref{tab:1}, the phase shifts for two
identical particles interacting via a Gaussian potential 
as a function of the relative kinetic energy. 
Results have been computed for a non-relativistic
and a semi-relativistic kinematics, and for two values of
the angular momentum $l$. As expected, phase shifts are similar for
low relative energy, and differ when energy increases.

\section{Summary}
\label{sec:summary}

The 3-dimensional Fourier grid Hamiltonian method, used in a previous
work to compute bound states \cite{brau97a}, appears as a convenient
method to find the scattering solutions of a
spinless Salpeter equation (or a Schr\"{o}dinger equation).
It has the advantage of simplicity since it requires only the
evaluation of the potential at some grid 
points and it generates directly the values of the radial part of the
wave function at the same grid points. Moreover, the method can be
extended to the cases of non-local interaction or coupled channel
equations. Up to our knowledge,
this is the first time that the scattering solutions of the spinless
Salpeter equation are presented.

\par Meson-meson scattering has been recently investigated in terms of
quark degrees of freedom within the framework of the non-relativistic
resonating group method \cite{ceul96}. From this work, it appears that
the use of a semi-relativistic kinematics is necessary to avoid
inconsistencies related to the non-relativistic formalism. We have
calculated a semi-relativistic version of the pion-pion scattering
equation. This equation is a scattering spinless Salpeter equation.
In this framework, the method presented here appears suitable to
calculate the
corresponding phase shifts \cite{ceul98}. 

\par The accuracy of the solutions of the numerical method presented
here can easily be controlled since it
depends only on two parameters: The 
number of grid points and the largest value of the radial distance 
considered to perform the calculation. This distance must be large
enough to fall in the region where the potential can be neglected
with respect to the asymptotic kinetic energy.
Both parameters can be automatically increased until a convergence 
is reached for phase shifts. 

\par The method involves the use of matrices of order 
$\left( N \times N \right)$, where $N$ is the number of grid
points. Generally, the most time consuming part of the method is the 
solution of the linear system. This is not a problem for 
modern computers, even for PC stations. Moreover, several powerful 
techniques exist and can be used at the best convenience. 
A demonstration program is available via anonymous FTP on 
{\tt umhsp02.umh.ac.be/pub/ftp\_pnt/}.

\acknowledgments
We thank Prof. R. Ceuleneer for useful discussions. 
                                                                                
\appendix
\section*{Particular case of the spinless Salpeter equation} 

The spinless Salpeter equation for two identical particles
interacting 
via a non-local potential can be written
\begin{equation}
\label{a1}
2 \sqrt{\vec p\,^2 + m^2}\, \Psi(\vec r\,)  
= M\, \Psi(\vec r\,)
- \int d\vec r\,'\, W(\vec r,\vec r\,')\, \Psi(\vec r\,').
\end{equation}
Acting on both sides with the square-root operator gives
\begin{equation}
\label{a2}
4 \left( \vec p\,^2 + m^2 \right) \Psi(\vec r\,)  
= M \left(
M \Psi(\vec r\,)
- \int d\vec r\,'\, W(\vec r,\vec r\,')\, \Psi(\vec r\,')
\right)
- 2 \int d\vec r\,'\, \sqrt{\vec p\,^2 + m^2}\,
W(\vec r,\vec r\,')\, \Psi(\vec r\,').
\end{equation}
If the potential has the form
$W(\vec r,\vec r\,') = V_0\, V(r)\, V(r')$,
and if we perform the integrations on angular variables, we obtain
\begin{equation}
\label{a4}
\left( \vec p\,^2 + m^2 - \frac{M^2}{4}\right) R_0(r) =
-\pi V_0 \left( M + 2 \sqrt{\vec p\,^2 + m^2} \right) V(r)
\int_0^\infty dr'\, r'^2\, V(r')\, R_0(r').
\end{equation}
where $R_0(r)$ is the radial part of the S-wave function
$\Psi(\vec r\,)$.
It has been shown in Ref.~\cite{brau97b} that
\begin{equation}
\label{a5}
\sqrt{\vec p\,^2 + m^2}\, e^{-mr} = \frac{4m}{\pi}\,K_0(mr),
\end{equation}
where $K_0(x)$ is a modified Bessel function \cite[p. 952]{grad80}.
In this
case, if we choose $V(r)=\exp(-mr)$, then Eq.~(\ref{a1}) reduces to
a non-local Schr\"{o}dinger-like equation
\begin{equation}
\label{a6}
\left( \frac{d^2}{dr^2} - m^2 + \frac{M^2}{4}\right) u_0(r) =
V_0 \left( M\pi e^{-mr} + 8 m K_0(mr) \right) r
\int_0^\infty dr'\, r'\, e^{-mr'}\, u_0(r'),
\end{equation}
where $u_0(r') = r' R_0(r')$.

\begin{table}
\protect\caption{Phase shifts for two identical particles with $m=1$
GeV as a function of the relative kinetic energy $E$. The interaction
is a Gaussian potential with $V_0=0.1$ GeV and $a=5$ GeV$^{-1}$ (see
Eq.~(\protect\ref{potg})). Results are given for a non-relativistic
(NR) and a semi-relativistic (SR) kinematics, and for two values of
the angular momentum $l$.}
\label{tab:1}
\begin{tabular}{cccc}
$l$ & $E$ (GeV) & $\delta_{\text{NR}}$ (rad) & $\delta_{\text{SR}}$
(rad) \\
\hline
0 & 0.001 & $-$0.192 & $-$0.189 \\
  & 0.01  & $-$0.627 & $-$0.619 \\
  & 0.1   &    1.363 &    1.376 \\
  & 1.    &    0.447 &    0.524 \\
  & 10.   &    0.156 &    0.256 \\
  &       &          &          \\
1 & 0.001 & $-$0.231 & $-$0.226 \\
  & 0.01  & $-$0.931 & $-$0.925 \\
  & 0.1   &    1.241 &    1.254 \\
  & 1.    &    0.479 &    0.517 \\
  & 10.   &    0.156 &    0.256 \\
\end{tabular}
\end{table}

\begin{thebibliography}{aa}
\bibitem{pres92} William H. Press, Saul A. Teukolsky, 
William T. Vetterling, and Brain P. Flannerey, {\em Numerical Recipes
in FORTRAN} (Cambridge University Press, 1992).
\bibitem{brau97a} F. Brau and C. Semay, J. Comput. Phys {\bf 139}, 127
(1998). 
\bibitem{mars89} C. Clay Marston and Gabriel G. Balint-Kurti, J.
Chem. Phys.
{\bf 91}, 3571 (1989).
\bibitem{bali91} Gabriel G. Balint-Kurti, Christopher L. Ward and 
C. Clay Marston, Comput. Phys. Commun. {\bf 67}, 285
(1991).
\bibitem{rodb67} Leonard S. Rodberg and Roy M. Thaler, {\em
Introduction to
the Quantum Theory of scattering} (Academic Press, New York, 1967).
\bibitem{ceul96} R. Ceuleneer, C. Semay and B. Silvestre-Brac, J. Phys.
G {\bf 22}, 1395 (1996). 
\bibitem{ceul98} R. Ceuleneer and C. Semay, ``Semi-relativistic RGM
calculations of pion-pion scattering'', in preparation. 
\bibitem{brau97b} F. Brau, {\em Integral equation formulation
of the spinless Salpeter equation}, Preprint Mons 1997, to appear in
J. Math. Phys. 
\bibitem{grad80} I.S. Gradshteyn and I.M. Ryzhik, {\em Tables of 
Integrals, Series, and Products} (Academic Press, New York, 1980).
\end{thebibliography}
\end{document}